\documentstyle[epsfig]{mn}

\title[The Norma Galaxy Cluster (ACO 3627): I.]
{The Norma Cluster (ACO 3627): I. A Dynamical Analysis of the Most Massive Cluster 
in the Great Attractor}
\author[Woudt et al.]
       {P.A. Woudt$^1$\thanks{E-mail: Patrick.Woudt@uct.ac.za}, 
        R.C. Kraan-Korteweg$^1$, J. Lucey$^2$, A.P. Fairall$^1$, S.A.W. Moore$^2$\\
        $^1$Department of Astronomy, University of Cape Town, Private Bag X3,
        Rondebosch 7701, South Africa\\
        $^2$Department of Physics, University of Durham, Durham DH1 3LE, United Kingdom}
\date{2007 May 22}

\begin{document}

\maketitle

\begin{abstract}
A detailed dynamical analysis of the nearby rich Norma cluster (ACO 3627) is presented. 
From radial velocities of 296 cluster members, we find a mean velocity of 4871 $\pm$ 54 km s$^{-1}$ and
a velocity dispersion of 925 km s$^{-1}$. The mean velocity of the E/S0 population
(4979 $\pm$ 85 km s$^{-1}$) is offset with respect to that of the S/Irr population (4812 $\pm$ 70 km s$^{-1}$)
by $\Delta v = 164$ km s$^{-1}$ in the cluster rest frame. This offset increases towards the core of the cluster. The E/S0 population is free
of any detectable substructure and appears relaxed. Its shape is clearly elongated with a position angle that is aligned along
the dominant large-scale structures in this region, the so-called Norma wall. The central cD galaxy has a very large 
peculiar velocity of 561 km s$^{-1}$ which is most probably related to an ongoing merger at the core of the cluster. 
The spiral/irregular galaxies reveal a large amount of substructure; two dynamically distinct subgroups within the overall 
spiral-population have been identified, located along the Norma wall elongation. The dynamical mass of the Norma cluster within its Abell radius is
$1 - 1.1 \times 10^{15} h^{-1}_{73}$ M$_{\odot}$. One of the cluster members, the spiral galaxy WKK\,6176 which recently was observed to have a 70 kpc 
X-ray tail, reveals numerous striking low-brightness filaments pointing away from the cluster centre suggesting strong interaction
with the intracluster medium.
\end{abstract}


\section{Introduction}
The observed velocity flow field of galaxies in the nearby Universe is largely dominated by the
Great Attractor (Dressler et al.~1987; Lynden-Bell et al.~1988; Tonry et al.~2000) and the $\sim$3 times more 
distant Shapley supercluster (Hudson et al.~2004). Both are extended 
overdensities in the large-scale mass distribution of the local Universe and both are thought to contribute
significantly to the peculiar motion of the Local Group (LG) (Lucey, Radburn-Smith \& Hudson 2005; 
Kocevski \& Ebeling 2006). The relative contribution of the Great Attractor and the 
Shapley supercluster to the motion of the LG, however, remains poorly determined (cf.~Erdo\u{g}du et al.~2006; 
Kocevski \& Ebeling 2006) and is still a matter of debate.

\begin{figure*}
\centerline{\hbox{\psfig{figure=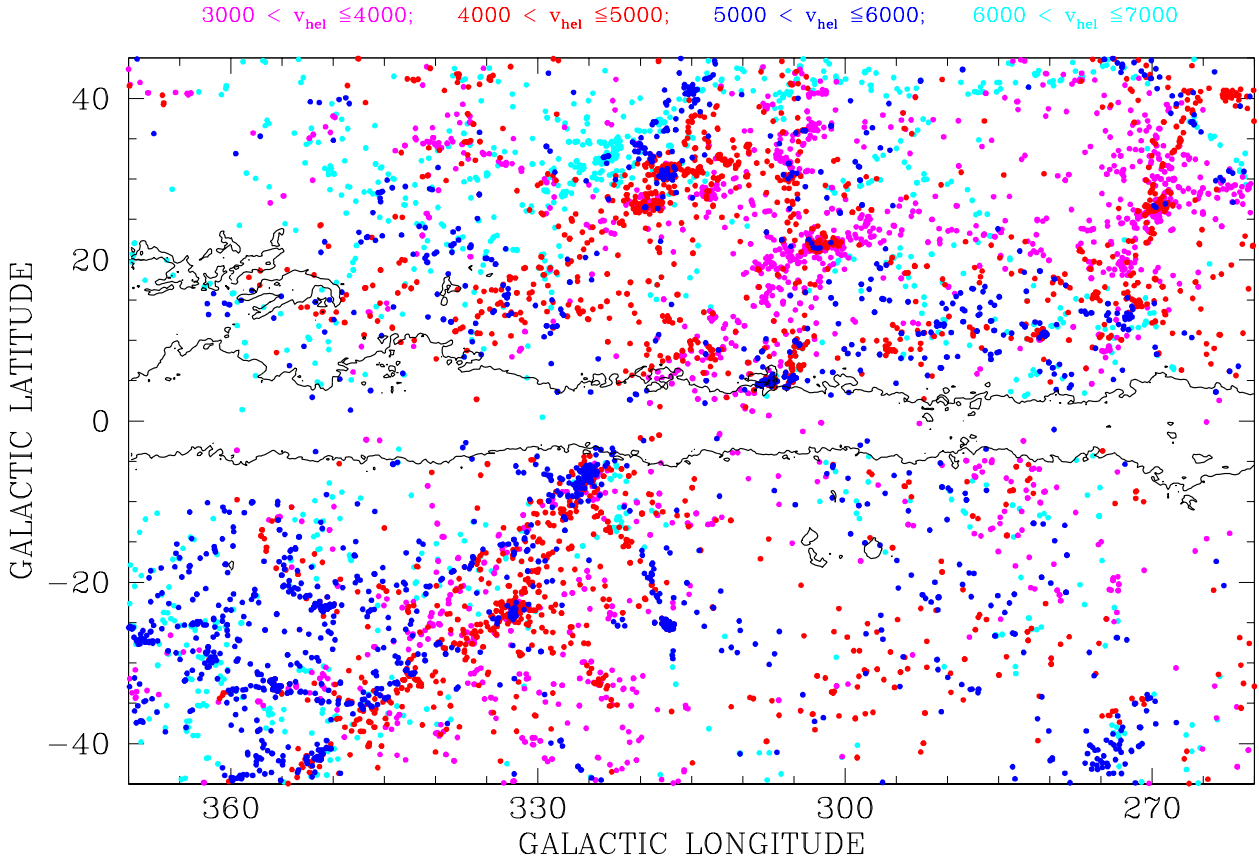,width=11.5cm,angle=-90}}}
 \caption{An overview of the large-scale structures in the Great Attractor region between 3000 $< v_{\rm hel} <$ 7000
km s$^{-1}$. The Norma cluster is located at ($\ell, b, v$) = ($325.3^{\circ}, -7.2^{\circ}, 4871$ km s$^{-1}$). Other
major clusters in this vista are the Pavo II cluster at ($\ell, b, v$) = ($332.3^{\circ}, -23.6^{\circ}, 4167$ km s$^{-1}$), 
the Centaurus cluster at ($\ell, b, v$) = ($302.4^{\circ}, +21.6^{\circ}, 3418$ km s$^{-1}$), the Hydra cluster at
($\ell, b, v$) = ($269.6^{\circ}, +26.5^{\circ}, 3777$ km s$^{-1}$), and the low-latitude
CIZA\,J1324.7--5736 and Cen-Crux clusters, at $(\ell, b, v$) = ($307.4^{\circ}, +5.0^{\circ}, 5700$ km s$^{-1}$) and
($305^{\circ}, +5^{\circ}, 6214$ km s$^{-1}$), respectively. The solid contour marks a line of equal Galactic foreground extinction
($A_B$ = 3$\fm0$, from Schlegel, Finkbeiner \& Davis 1998). }
 \label{lssoverview}
\end{figure*}

The Shapley supercluster (SCL) is clearly visible as an overdensity in the 
distribution of Abell clusters (Scaramella et al.~1989; Einasto et al.~1997; Proust et al.~2006), 
whereas the Great Attractor (GA) and its location was identified first and foremost from the systematic
peculiar velocities of galaxies streaming towards this apex (e.g. Lynden-Bell et al.~1988), and confirmed
later from reconstructed mass-density fields of the local Universe (Dekel 1994; Kolatt, Dekel \& Lahav~1995; 
Erdo\u{g}du et al.~2006). However, no significant overdensity was obvious in the distribution of galaxies or Abell clusters
at the position of the GA (Lynden-Bell \& Lahav 1988). 
This is not surprising given the location of the GA with respect to the
Zone of Avoidance (ZOA). Kolatt et al.~(1995) locate the approximate centre of the extended GA overdensity at 
($\ell, b, v$) $\sim$ (320$^{\circ}$, 0$^{\circ}$, 4000 km s$^{-1}$) based on their smoothed reconstructed 
mass-density field. The uncertainty in this position is $\sim 17^{\circ}$ as a result of the applied smoothing 
of 1200 km s$^{-1}$; this smoothing is necessary given the sparsely sampled data (Kolatt et al.~1995).
Towards such low Galactic latitudes,
the view of the extragalactic light distribution is increasingly reduced by the dust and stars in the 
Milky Way. As a result, a large part of the GA overdensity is hidden from view by the Milky Way and
early attempts to quantify the nature and extent of the GA (e.g. Dressler 1988; Hudson 1993a, 
1993b; Rowan-Robinson et al.~1990; Jahoda \& Mushotsky 1989) were unsatisfactory: the clear
(and significant) mismatch between the inferred mass of the GA and the visible galaxy distribution
could not be understood. 

A deep optical galaxy search at low Galactic latitudes in the GA region (Woudt \& Kraan-Korteweg 2001) has
lifted part of the veil of the Milky Way. Close to the predicted centre of the GA, the Norma cluster (ACO 3627:
Abell, Corwin \& Olowin 1989) has been identified as the most massive cluster in the GA region 
(Kraan-Korteweg et al.~1996; Woudt 1998). Abell et al.~(1989) classified this cluster as an irregular (I) cluster
with Bautz-Morgan type I (Bautz \& Morgan 1970). They furthermore classify it as a richness-class 1 cluster with 
`59?' galaxies in the magnitude interval $m_3$ and $m_3$ + 2, where $m_3$ corresponds to the magnitude of the 
third brightest galaxy in the cluster. Independently, 
X-ray observations of the Norma cluster from ROSAT (B\"ohringer et al.~1996) and ASCA (Tamura et al.~1998)
confirm the massive nature of this cluster. From our deep optical galaxy survey in the ZOA in the 
general GA region, and our follow-up redshift survey (Fairall, Woudt \& Kraan-Korteweg 1998;
Woudt, Kraan-Korteweg \& Fairall 1999; Woudt et al.~2004), a clearer view of the obscured 
GA overdensity has emerged. The Norma cluster is the central cluster in a web of 
connected filaments and wall-like structures (Woudt 1998; Kraan-Korteweg \& Lahav 2000; Radburn-Smith et al.~2006), analogous
to the structures observed in and around major mass concentrations in the $\Lambda$-CDM (cold dark matter) 
Millenium simulation (Springel et al.~2005).

One of the most prominent newly identified structures is a great-wall-like structure with the Norma cluster at its centre
which we dubbed the Norma supercluster (Woudt 1998; Fairall et al.~1998),
a wall of galaxies which runs nearly parallel to the Galactic Plane (Kraan-Korteweg \& Lahav 2000; 
Radburn-Smith et al.~2006) connecting the Pavo II cluster
with the Norma cluster and continuing across (and nearly parallel to) the Galactic Plane to the more distant Vela overdensity
(Kraan-Korteweg, Fairall \& Balkowski 1995) via the Cen-Crux cluster (Woudt 1998). Fig.~\ref{lssoverview} gives a clear overview
of the dominant large-scale structures in the Great Attractor region.

Support for the prominence of the Norma SCL
has come from various complementary multiwavelength studies at lower Galactic latitudes
such as the detection of several further clusters embedded in the Norma SCL.
An X-ray search for highly obscured clusters in the ZOA (Ebeling, Mullis \& Tully~2002) revealed 
the second most massive cluster in the Norma SCL, namely CIZA\,J1324.7--5736. This cluster is $\sim$50\%--70\% less massive
than the Norma cluster (Radburn-Smith et al.~2006) and is located at $(\ell, b, v) \sim (307.4^{\circ}, +5.0^{\circ}, 5700$ km
s$^{-1}$). Deep near-infrared observations (Nagayama et al.~2004) furthermore uncovered a low-mass cluster around PKS\,1343-601 at 
$(\ell, b, v) \sim (309.7^{\circ}, +1.7^{\circ}, 3900$ km s$^{-1}$), also within the Norma SCL.
Apart from this significant collection of clusters, a general overdensity along the Norma SCL
is also clearly present in the Parkes deep H\,I multibeam ZOA survey (Kraan-Korteweg et al.~2005). 

\begin{figure}
\centerline{\hbox{\psfig{figure=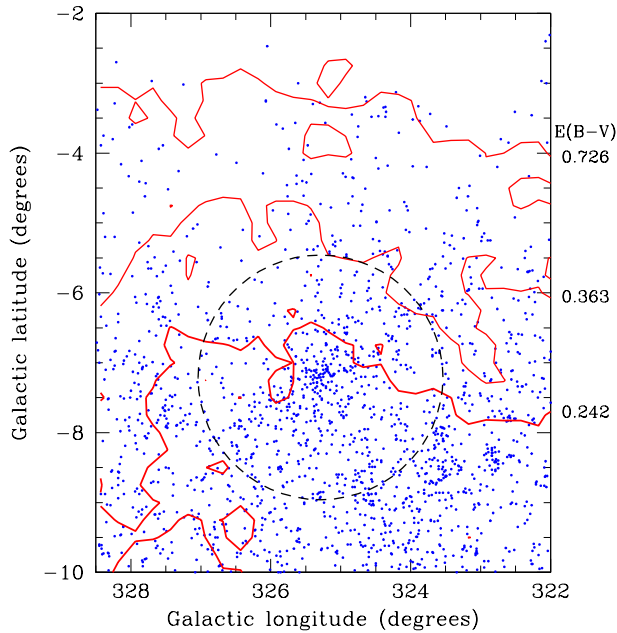,width=8.2cm}}}
 \caption{The distribution in Galactic coordinates of optically-detected galaxies (Woudt \& Kraan-Korteweg 2001)
around the Norma cluster. The contours indicate lines of equal Galactic reddening 
(from the DIRBE/IRAS reddening maps, Schlegel et al.~1998) of $E(B-V)$ = 0.242, 0.363 and
0.726 mag, respectively. Assuming a standard Galactic reddening law (Cardelli et al.~1989), 
these values correspond to $A_{\rm B}$ = 1.0, 1.5 and 3.0 mag, respectively. The dashed circle
marks the Abell radius of the Norma cluster.}
 \label{distribution}
\end{figure}

In the first of a series of papers investigating the Norma cluster, we present
a detailed dynamical analysis of this cluster, the most massive cluster in the Great Attractor overdensity,
centrally located in a cosmic web of filaments and wall-like structures.

Figure~\ref{distribution} shows the distribution of the optically-detected galaxies 
(Woudt \& Kraan-Korteweg 2001) in the general direction of the Norma cluster, where the 
Abell radius ($R_A \equiv 1\farcm7/z$) of the Norma cluster is indicated by the dashed circle. 
At the redshift of the Norma cluster (see Sect.~3), the Abell radius corresponds to an
angular radius of 1.75$^{\circ}$. Assuming a Hubble constant of $H_0 = 73$ km s$^{-1}$ Mpc$^{-1}$ and
the cosmological concordance model (assumed throughout this paper), the Abell radius 
corresponds to a physical size of 2.0 Mpc (the cosmology-corrected angular scale at this distance is 1.16 Mpc per
degree).
Contours of equal Galactic foreground extinction, taken from the DIRBE/IRAS Galactic reddening map 
(Schlegel, Finkbeiner \& Davis~1998), are overlayed on the galaxy distribution in Fig.~\ref{distribution} 
($A_{\rm B}$ = 1.0, 1.5 and 3.0 mag, respectively; Cardelli, Clayton \& Mathis~1989) and show that 
the Galactic foreground extinction within the Abell radius of the Norma cluster 
is moderate, $A_{\rm B} \le 1.5$ mag. 

Within the Abell radius, there are 603 optically-detected galaxies with observed diameters in excess of 12$''$
(Woudt \& Kraan-Korteweg 2001) and 219 (near-infrared-detected) galaxies in the extended source catalogue (XSC)
of the 2 Micron All-Sky Survey (2MASS, Skrutskie et al.~2006). The 2MASS galaxies in the Norma cluster represent a subset
of the 603 optically-detected galaxies, although not all the 2MASS galaxies have an optical counterpart; 165 of the 219 2MASS 
galaxies (75\%) were also found by Woudt \& Kraan-Korteweg (2001). For the brighter 2MASS galaxies (10$''$-aperture $K_s$-band 
$<$ 12.5 mag), the overlap between 2MASS and the optical survey is excellent: 97\% of the 2MASS galaxies have an optical counterpart.
It should be noted that at the position of the Norma cluster ($\ell, b \sim 325^{\circ}, -7^{\circ}$)
star-crowding is the primary limiting factor, not the Galactic foreground extinction. The star-crowding leaves 
a Zone of Avoidance imprint on the 2MASS XSC catalogue near the Galactic Bulge (Kraan-Korteweg \& Jarrett 2005)
and the Norma cluster is located on the edge of this Zone of Avoidance. At moderate extinction ($A_B \le 3$ mag), 
but in the presence of severe star-crowding, optical surveys still retrieve the most complete galaxy distribution 
in the Zone of Avoidance (Kraan-Korteweg \& Jarrett 2005). 

We have obtainted 129 new redshifts of galaxies within the Abell radius of the Norma cluster using the 2dF spectrograph
at the Anglo-Australian Observatory. These new observations are presented in Section 2. In Section 3 all the redshifts 
obtained to date are combined and a detailed dynamical analysis of the cluster based on 296 cluster members is presented. 
In Section 4, we discuss a few
individual galaxies in the Norma cluster of dynamical interest.

\section{2dF spectroscopy} 

Spectra were obtained with the 2dF facility (Lewis et al.~2002) on the 3.9m Anglo-Australian Telescope.
Full details of the observing 2dF setup used for observations are given in
Table~\ref{2dfsetup}. As the main objective was to measure the velocity
disperisons of the cluster's early-type galaxies the
1200V gratings were used in each of the 2dF spectrographs.
These gave a FWHM resolution of $\sim$ 125 km s$^{-1}$ at Mg\,$b$ which is
sufficient to determine velocity dispersions down
to $\sim$ 60 km s$^{-1}$. In all, three fibre configurations were observed.
Spectra were extracted from the raw data frames,
wavelength calibrated and sky-subtracted using
the AAO 2dfdr software package\footnote{\tt http://www.aao.gov.au/2df/software.html\#2dfdr}.
Redshifts were determined via cross-correlation for
the absorption line spectra and/or the direct measurement of
emission lines.

\begin{table}
 \centering
  \caption{2dF setup used.}
  \begin{tabular}{@{}ll@{}}
\hline
Date of Observations           & 2001 May 30 \\
Field centre (J2000.0)         & $16^h15^m01.8^s$  $-60^{\circ}54'24''$ \\
Number of fibre config. &   3  \\
Total exposure times           & $5 \times 1200$ s, $5 \times 1200$ s, $4 \times 1200$ s \\
Fibre size             	 &	2.1 arcsec (= 0.68 kpc at Norma) \\
Grating              	  &	1200V \\
Wavelength coverage       &       4700 -- 5840 {\AA} \\
Resolution (FWHM) 	&	2.2 {\AA} \\
Wavelength pixel scale   &        1.1 {\AA} \\
\hline
\end{tabular}
\label{2dfsetup}
\end{table}

The 2dF spectroscopic observations focussed on the determination of
accurate velocity dispersions of early-type galaxies in the Norma cluster for a
Fundamental Plane analysis of the cluster. The primary target list therefore consisted of known bright 
ellipticals in the cluster (Woudt \& Kraan-Korteweg 2001). However, we used the spare fibres of the 2dF 
spectrograph to extend the redshift coverage of the Norma cluster. Galaxies were primarily 
selected from the optical catalogue of Woudt \& Kraan-Korteweg (2001) and the 2MASS XSC, indicated by
`WKK' and `2MASX\,J', respectively in Table~\ref{2dftable}. Additional galaxies were identified on deep
$R_C$ images taken with the ESO/MPG 2.2-m telescope and the Wide Field Imager (see Sect.~4). These are
identified as `ZOA\,J' in Table~\ref{2dftable}.

Redshifts were obtained for 182 galaxies, 53 of which had a previous measurement. 
For 76 galaxies, multiple measurements were obtained to gauge the internal accuracy of
the 2dF spectrograph. Table~\ref{2dftable} shows a representative sample of the results obtained
from the 2dF spectroscopy. The full table is available online.

\begin{table*}
 \centering
  \caption{A representative sample of the results of the 2dF spectroscopy.}
  \begin{tabular}{@{}lcccccc@{}} 
\hline
 Identification$^*$  & {RA (2000.0)} &  {DEC (2000.0)} & {\it v$_{\rm abs}$} & {\it v$_{\rm em}$} \\ 
                 &               &                 & (km s$^{-1}$)       & (km s$^{-1}$) \\ 
\hline
 ZOA\,J16070347-6113587   & 16 07 03.465  &--61 13 58.74 & \hfill 15868 & \hfill        \\
 WKK\,5916                & 16 07 50.369  &--61 10 06.84 & \hfill  3053 & \hfill        \\
 WKK\,5920                & 16 07 52.618  &--60 31 12.95 & \hfill  4762 & \hfill        \\
 WKK\,5926                & 16 08 08.744  &--61 12 44.37 & \hfill       & \hfill  15856 \\
 ZOA\,J16081355-6109377   & 16 08 13.548  &--61 09 37.65 & \hfill  3640 & \hfill        \\
 2MASX\,J16082135-6044498 & 16 08 21.312  &--60 44 50.20 & \hfill 29813 & \hfill        \\
 ZOA\,J16083012-6039511   & 16 08 30.118  &--60 39 51.08 & \hfill  6040 & \hfill        \\
 WKK\,5958                & 16 09 01.326  &--60 52 03.70 & \hfill 15708 & \hfill        \\
 WKK\,5964                & 16 09 06.402  &--60 59 07.70 & \hfill  4711 & \hfill        \\
 ZOA\,J16091138-6108285   & 16 09 11.377  &--61 08 28.45 & \hfill       & \hfill  15674 \\
\hline
\end{tabular}
\label{2dftable}
{\newline \footnotesize{$^*$ In column 1, the WKK identification (Woudt \& Kraan-Korteweg 2001) is given if the galaxy has been identified by 
WKK. \hfill  \ 
\newline Alternative names for galaxies are given if a WKK identification was unavailable but when the galaxy has been identified already 
in \hfill \ 
\newline another survey (e.g.~`2MASX\,J', Skrutskie et al.~2006). A new identification was given to galaxies not yet catalogued in the 
literature,\hfill  \ 
\newline but which we identified from deep $R_C$ imaging (e.g.~`ZOA\,J', see Sect.~4). \hfill }}
\end{table*}

\begin{figure}
\centerline{\hbox{\psfig{figure=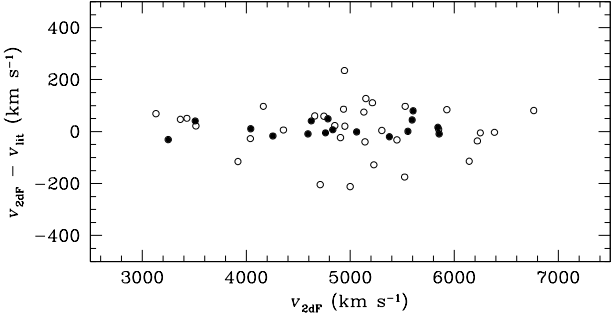,width=8.2cm}}}
 \caption{A comparison of the 2dF velocities with previously determined velocities.}
 \label{2dfcompext}
\end{figure}

Figure~\ref{2dfcompext} shows a comparison of the measured 2dF heliocentric velocities with
measurements from the literature. The vast majority of these previous measurements were obtained in the course
of our ZOA redshift survey (SAAO: Woudt et al.~1999; MEFOS: Woudt et al.~2004). The overall agreement is very good:
$$v_{\rm 2dF} - v_{\rm lit} = -6 \pm 17 \  {\rm km \, s^{-1}}$$
with a dispersion of $\sigma_{\rm ext, all}$ = 124 km s$^{-1}$ (based on 51 galaxies). 
Only one galaxy revealed a discrepent heliocentric velocity; for
WKK\,6329, the 2dF spectroscopy resulted in $v = 4749 \pm 35$ km s$^{-1}$ as compared to the previously low 
signal-to-noise value for this galaxy of 2477 $\pm$ 250 km s$^{-1}$ (Woudt et al.~1999).

We then compared the 2dF results with a subset of the literature sample, namely those for which redshifts
were obtained with the MEFOS multi-fibre spectrograph (Woudt et al.~2004). This subset has the
most accurate redshifts available for the Norma cluster. There are 16 galaxies in common between 
2dF and MEFOS (the filled circles in Fig.~\ref{2dfcompext}). The 
agreement is again excellent, with a lower rms ($\sigma_{\rm ext, MEFOS}$ = 31 km s$^{-1}$) than the previous
comparison (which included the SAAO measurements),
$$v_{\rm 2dF} - v_{\rm MEFOS} = +12 \pm 8 \ {\rm km \, s^{-1}}.$$

Given the primary goal of obtaining accurate velocity dispersions from the 2dF spectroscopy,
we have observed a large number of galaxies repeatedly to gauge the internal 
uncertainty: 69 galaxies were observed twice and 7 galaxies had three independent velocity measurements.
For these repeated observations we find $\sigma_{\rm int}$ = 33 km s$^{-1}$ over the entire
range of observed velocities. This is comparable to the external comparison with the MEFOS spectroscopy.
Based on these independent evaluations, we have assigned a standard error of 35 km s$^{-1}$ to each of the 2dF
velocities.

\section{Dynamical Analysis}

\subsection{Cluster membership}

\begin{figure}
\centerline{\hbox{\psfig{figure=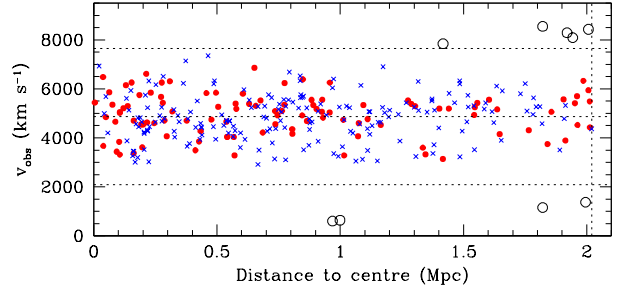,width=8.2cm}}}
 \caption{Galaxies with known redshifts as a function of distance to the central cD galaxies (WKK\,6269) in
the Norma cluster. The Abell radius ($R_A$) is indicated by the vertical dotted line, were the velocity centroid
and the upper and lower 3$\sigma$ limits are indicated by the horizontal dotted line. 
The E/S0 galaxies in the Norma cluster are plotted as filled circles and the S/Irr galaxies
in the Norma cluster are shown as crosses. Galaxies deemed non-members are indicated by the open circles.}
 \label{raddist}
\end{figure}

\begin{figure}
\centerline{\hbox{\psfig{figure=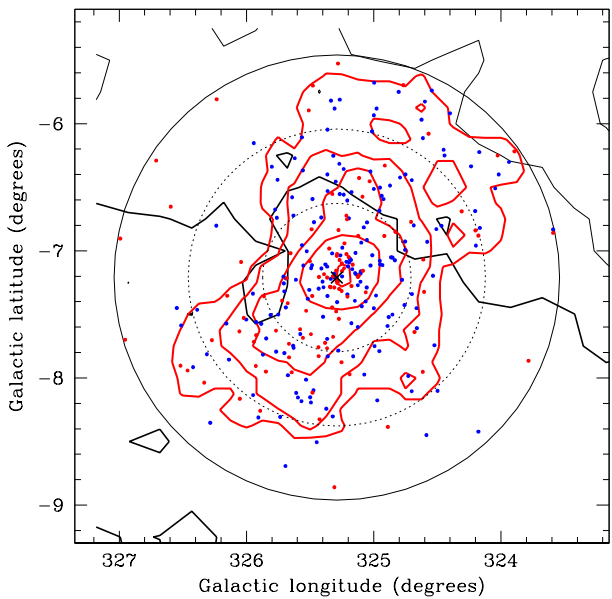,width=8.2cm}}}
 \caption{Galaxy density contours determined from the distribution (in Galactic coordinates) of the 296 likely cluster 
members. The contours have been normalised by the total number of galaxies in the sample ($N_s$); they correspond to 
0.063 $N_s$, 0.125 $N_s$, 0.25 $N_s$, 0.50 $N_s$ and 1.0 $N_s$ galaxies per square degree, respectively. The location
of WKK\,6269 is indicated by the central cross, and concentric cluster radii of $\frac{1}{3} R_A$, $\frac{2}{3} R_A$ and
$R_A$ are shown as the two dashed circles and the solid circle, respectively. The extinction contours are as in 
Fig.~\ref{distribution}.}
 \label{density}
\end{figure}

With the new 2dF observations described above, radial velocities are now available for 305 galaxies 
within the Abell radius of the Norma cluster for the velocity range 0 -- 9500 km s$^{-1}$.
The velocity distribution of these galaxies as a function of distance to the Norma cluster centre
is shown in Fig.~\ref{raddist}. The centre of the cluster was taken as the cD galaxy WKK\,6269, a strong 
wide-angle-tail radio-continuum source (Jones \& McAdam 1992, 1996) located at the peak in the 0.7--10 keV 
ASCA map of the Norma cluster (figure 1 of Tamura et al.~1998). The velocity centroid (see Sect.~3.2.3)
of the Norma cluster is 4871 km s$^{-1}$ and is marked by the central horizontal dashed line. The velocity limits
for cluster membership are taken as $\pm$3 times the velocity scale/dispersion (925 km s$^{-1}$) around the velocity centroid;
these limits are shown as the upper and lower horizontal dashed line in Fig.~\ref{raddist}.

Nine galaxies are distinct outliers (open circles in Fig.~\ref{raddist}) and have been rejected from our
subsequent analysis. This leaves 296 likely cluster members, of which 107 have been classified as elliptical or
lenticular (E/S0: filled circles in Fig.~\ref{raddist}) and 189 are either spirals or irregulars
(S/Irr: crosses in Fig.~\ref{raddist}) (Woudt \& Kraan-Korteweg 2001).

The galaxy density contours determined from the 296 cluster members, displayed in Fig.~\ref{density}, show that the cluster
is strongly elongated along a position angle which is aligned with the Norma wall (compare Figs.~\ref{lssoverview} 
and \ref{density}).
Since the elongation is nearly perpendicular to the Galactic extinction contours, it seems very unlikely that selective
extinction effects are the cause of the observed elongation. The peak
of the galaxy-density distribution is located at right ascension and declination $16^h14^m42^s$, $-60^{\circ}55'52''$ (J2000.0), 
about 3 arcmin from WKK\,6269 at $16^h15^m03.6^s$, $-60^{\circ}54'26''$ (J2000.0), our adopted centre.

\subsection{Substructure statistics}

We employed the statistical tests described by Pinkney et al.~(1993, 1996)
in analysing the dynamical structure of the Norma cluster. This array of statistical tools consists of 
one-dimensional tests (analysing the shape of the velocity histogram), two-dimensional tests 
(checking for substructure in the on-sky distribution), and three-dimensional tests (using
velocity and positional information). Among the latter, the Dressler-Shectman (DS) $\delta$-test
(Dressler \& Shectman 1988) is a particularly powerful and frequently used method to quantify substructure 
(e.g.~Pinkney et al.~1993; Oegerle \& Hill 2001; Pimbblet, Roseboom \& Doyle 2006).
This test calculates the mean velocity ($\langle v \rangle_{\rm local}$)
and the standard deviation ($\sigma_{\rm local}$) for each galaxy and its $N_{nn}$ (= $\sqrt{N}$) nearest 
neighbours, where $N$ represents the total number of galaxies in the sample; often only the 10 nearest galaxies 
are used in this analysis. These local parameters are then
compared with the global mean ($\langle v \rangle$) and standard deviation ($\sigma$)
of all the galaxies in the sample. For each galaxy, $\delta_i$ is calculated where $\delta_i$ is given by
$$\delta_i^2 = \left( \frac{N_{nn} + 1}{\sigma^2} \right) \left[(\langle v \rangle_{\rm local} - 
\langle v \rangle)^2 + (\sigma_{\rm local} - \sigma)^2 \right].$$
The cumulative deviation $\Delta$ is defined as the sum of all $\delta_i$'s. If no subclustering is
present, $\Delta$ is approximately equal to the number of galaxies in the sample ($N$).

In the following subsections we analyse the Norma cluster at three incremental radii, starting with the inner
core of the cluster ($R < 0.67$ Mpc), double this radius ($R < 1.35$ Mpc) and three times this radius 
out to the Abell radius ($R < 2.02$ Mpc).

\begin{figure*}
\centerline{\hbox{\psfig{figure=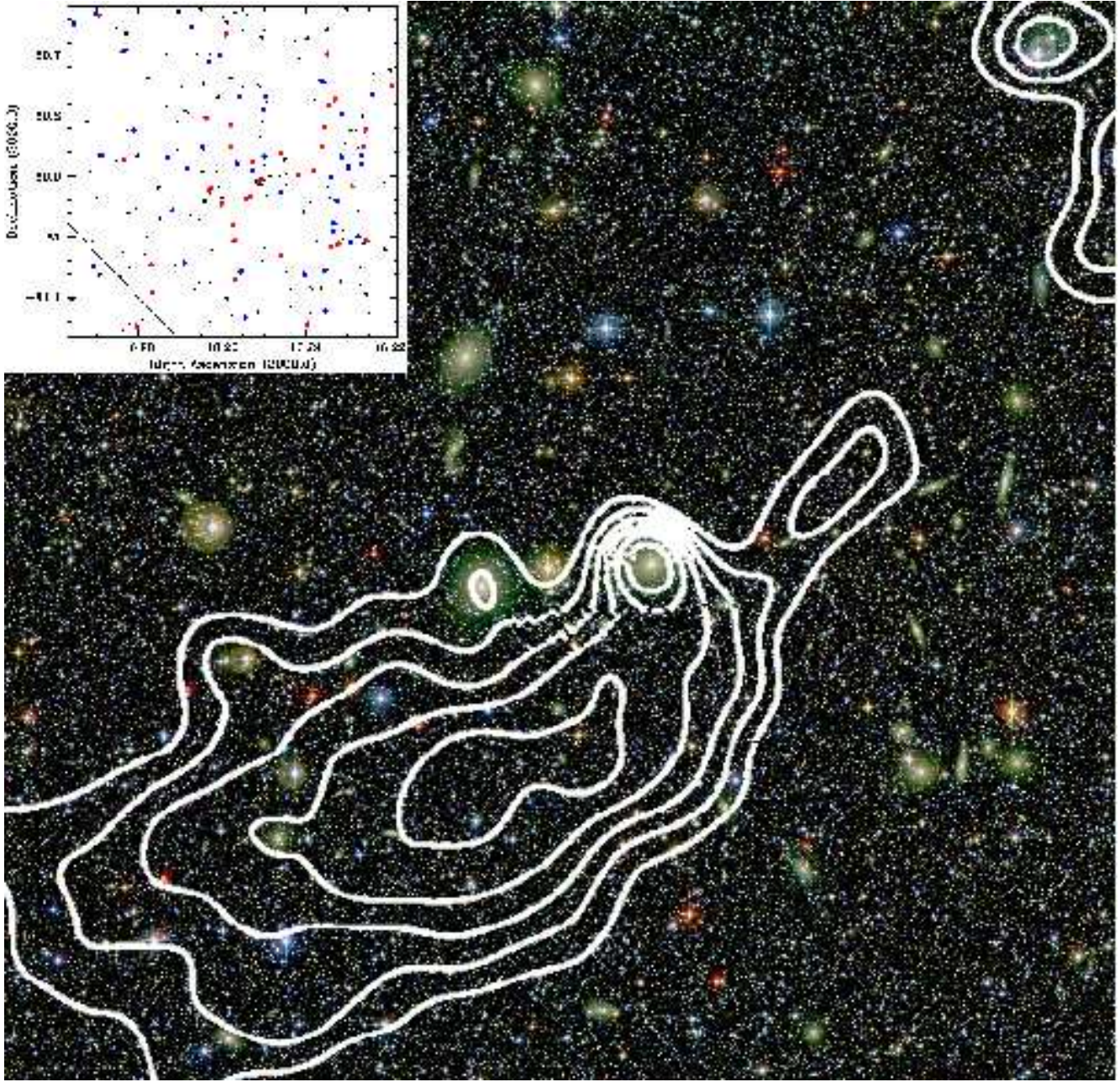,width=17.6cm}}}
 \caption{An optical colour image of the central $\sim$0.66 Mpc $\times$ 0.66 Mpc of the Norma cluster. The white contours
show the X-ray subgroup identified from ROSAT observations (reproduced from B\"ohringer et al.~1996) overlayed on the optical
galaxy distribution. The central cD galaxy (WKK\,6269) coincides with the peak in the X-ray emission. The inset shows the 
distribution of the identified galaxies in this field of view in equatorial coordinates (E/S0 cluster members: red dots, 
S/Irr cluster members: blue dots, WKK\,galaxies with no redshift information: large black dots, 
small galaxies identified from the Wide Field Image data: small black dots). For reference when comparing this 
figure with the distribution in Galactic coordinates, a line of equal
Galactic latitude ($b = -7.5^{\circ}$) is drawn as a diagonal dashed line in the inset.}
 \label{optxray}
\end{figure*}

\subsubsection{The inner $\frac{1}{3}$ Abell radius (= 0.67 Mpc)}

Figure~\ref{optxray} shows the optical image of the central $34 \times 34$ arcmin of the Norma cluster,
obtained with the 2.2-m MPG/ESO telescope at la Silla and the Wide Field Imager during three nights in 
1999 May (see also Section 4). The area displayed in Fig.~\ref{optxray} corresponds to $\sim$1.6 times 
the core radius ($R_{\rm c}$; King 1966) of the Norma cluster, where $R_{\rm c, \, opt}$ = 10$\farcm$4
$\pm$ 1$\farcm$1 (optical: Kraan-Korteweg et al.~1996) and $R_{\rm c,\, X}$ = 9$\farcm$95 $\pm$ 1$\farcm$0 
(X-ray; B\"ohringer et al.~1996) for the Norma cluster. In terms of the 
Abell radius, Fig.~\ref{optxray} displays the inner $\sim \frac{1}{6}$ $R_A$.


Superimposed on the optical colour image are the contours of the X-ray subcluster identified by B\"ohringer
et al.~(1996) (reproduced from their figure 2). Note that these contours mark the subcluster only and that the main 
cluster has been subtracted as described in B\"ohringer et al.~(1996). The inset in Fig.~\ref{optxray} shows
the corresponding sky distribution (in equatorial coordinates) of the identified galaxies in this field of view. 
The red and blue dots are confirmed cluster members, where
the red dots mark E/S0 galaxies, and the blue dots correspond to S/Irr galaxies. The large black dots are 
galaxies (without a redshift) identified in our deep optical survey (Woudt \& Kraan-Korteweg 2001) and the 
small black dots are galaxies (also without a redshift) identified on the deep $R_C$-band images taken 
with the Wide Field Imager. The central cD galaxy (WKK\,6269, see the discussion in Sect.~3.4 on the peculiar velocity
of this galaxy) is indicated by the black-encircled red dot.

\begin{figure}
\centerline{\hbox{\psfig{figure=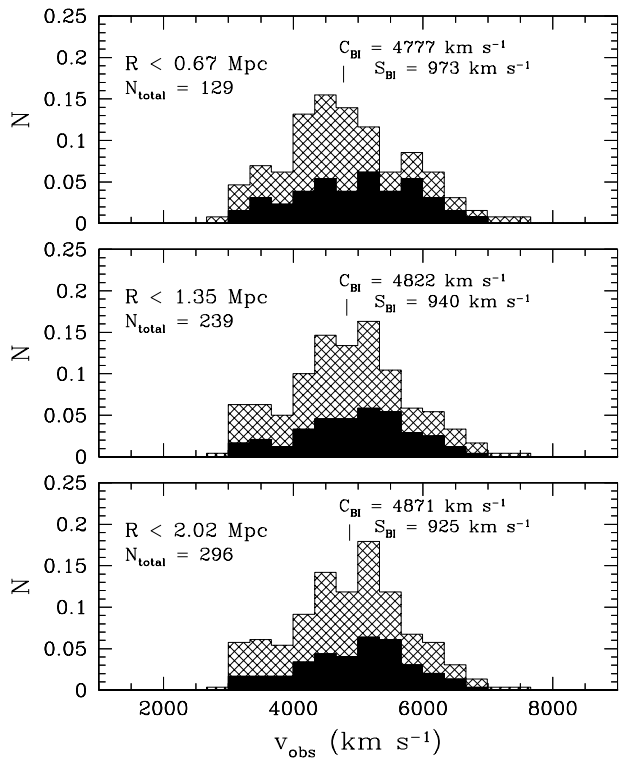,width=8.2cm}}}
 \caption{The normalised velocity distribution of cluster members (cross-hatched histogram)
within $R < 0.67$ Mpc (upper panel), $R < 1.35$ Mpc (middle panel) and $R < 2.02$ Mpc (lower panel). In each
of the panels, the velocity distribution of the E/S0 population is shown by the solid histogram.}
 \label{vhist}
\end{figure}

Within $R \le \frac{1}{3} R_A$, there are 129 galaxies confirmed as cluster members; 53 galaxies are E/S0 (41\%) and 76 
have been classified as S/Irr (59\%), respectively. 
The velocity histogram of the 129 cluster members -- shown in the upper panel of Fig.~\ref{vhist} as
the cross-hatched distribution -- is consistent with being Gaussian and has a biweight 
velocity centroid  ($C_{\rm BI}$) and scale ($S_{\rm BI}$) (Beers, Flynn \& Gebbhardt 1990) 
of 4777 $\pm$ 86 km s$^{-1}$ and 973 km s$^{-1}$, respectively. When separating the sample into the
elliptical and spiral population, there are some marked differences between these galaxy populations.
Firstly, the biweight velocity centroid of the two populations differ by 334 km s$^{-1}$ 
(see Table~{\ref{norsubstr}). The statistical significance of this difference -- by analogy to the arguments used in 
the peculiar velocity discussion (Sect.~3.4) -- is $S_{\rm V} = 1.9$.
Secondly, contrary to the velocity distribution of the elliptical galaxies 
(the dark shaded histogram in the top panel of Fig.~\ref{vhist}), the velocity distribution of the 
spiral galaxies is non-Gaussian with hints of skewness and kurtosis.

In addition, the elliptical population is strongly elongated along a position
angle in the equatorial on-sky projection of 116$^{\circ}$ (measured counter-clockwise from North),
whereas the distribution of the spiral galaxies is largely spherical. The position angle of 116$^{\circ}$ in
equatorial coordinates corresponds to a position angle of 160$^{\circ}$ (measured counter-clockwise from
North) in the galactic-coordinate distribution at the position of the Norma cluster. The position angle of the elongated
distribution is indicated by the arrows in the top-left panel of Fig.~\ref{dsall}.
The location
of the X-ray subcluster is indicated by the solid black line in the left panels of Fig.~\ref{dsall}. 
It is interesting to note the approximate alignment of the X-ray subgroup with the 
elongated distribution of the E/S0 galaxies. 
Statistical significance of the two- and three-dimensional substructure tests
are calculated by means of Monte Carlo (MC) simulations (Pinkney et al.~1996). The observed sample is compared
to 500 simulated samples. In the case of the elongation of the elliptical population, only 4 out of the 500 
simulations showed a larger degree of elongation.

Despite the above mentioned differences between the elliptical and spiral galaxy population, and despite the
presence of the X-ray subcluster, the Dressler-Shectman $\delta$-test showed no clear sign of substructure. The
combined sample, as well as the E/S0 galaxy sample (and to a lesser extent the S/Irr galaxies) are formally consistent with no 
substructure.  The $\Delta$ values for the combined sample and the individual galaxy populations are given 
in Table~\ref{norsubstr}, 
together with the average value of $\Delta$ after 500 Monte Carlo simulations. 

\begin{table*}
 \centering
  \caption{An investigation into substructuring in the Norma cluster.}
  \begin{tabular}{@{}lcccc@{}} 
\hline
  &   & $\frac{1}{3} R_A$ (= 0.67 Mpc) &  $\frac{2}{3} R_A$ (= 1.35 Mpc)  & {$1 R_A$ (= 2.02 Mpc)} \\
\hline
 & & \multicolumn{3}{c}{\underline{All galaxies}}\\
$C_{\rm BI}$   & (km s$^{-1}$)    &  4777 $\pm$ 86 &  4822 $\pm$ 61 &  4871 $\pm$ 54 \\ 
$S_{\rm BI}$  & (km s$^{-1}$)    &  973           &  940           &   925         \\
$N$  &                  &  129           &  239           &   296          \\
$\Delta^*$                      &   &  135.7 (38\%)  &  262.7 (23\%)    &   353.1  (4\%)  \\
$\langle {\Delta} \rangle_{500}$  &   &  131.4      &  244.0      &   301.4      \\
$v_{\rm pec}$ cD & (km s$^{-1}$)   &  653         &  609        &   561          \\
$S_{\rm V}$         &              & 6.5          &  7.6        &   7.5         \\
 & & \multicolumn{3}{c}{\underline{Elliptical and lenticular galaxies}}\\
$C_{\rm BI}$   & (km s$^{-1}$)    &  4951 $\pm$ 132 &  4962 $\pm$ 97 &  4979 $\pm$ 85 \\
$S_{\rm BI}$   & (km s$^{-1}$)    &  964            &  901           &   877         \\
$N$  &                  &  53             &  86            &   107          \\
$\Delta^*$                      &   &   44.8 (75\%)  &  83.7 (52\%)    &   107.7  (47\%)  \\
$\langle {\Delta} \rangle_{500}$  &   &  49.6      &  84.4      &   106.9      \\
 & & \multicolumn{3}{c}{\underline{Spiral and irregular galaxies}}\\
$C_{\rm BI}$   & (km s$^{-1}$)    &  4617 $\pm$ 109 &  4740 $\pm$ 78 &  4812 $\pm$ 70 \\
$S_{\rm BI}$   & (km s$^{-1}$)    &  949            &  965           &   957         \\
$N$  &                  &  76             &  153           &   189          \\
$\Delta^*$                       &   &  98.2 (14\%)  &  186.3 (8\%)    &   247.4  (1\%)  \\
$\langle {\Delta} \rangle_{500}$   &   &  85.6      &  158.0      &   191.2      \\
\hline
\end{tabular}
\label{norsubstr}
{\newline \footnotesize{$^*$ The percentages given after each value of $\Delta$ reflect the percentage of Monte Carlo simulations which showed
a higher amount of \hfill \ 
\newline subclustering than the actual observed sample. Percentages below 10\% indicate significant subclustering.\hfill \ }}
\end{table*}

\begin{figure*}
\centerline{\hbox{\psfig{figure=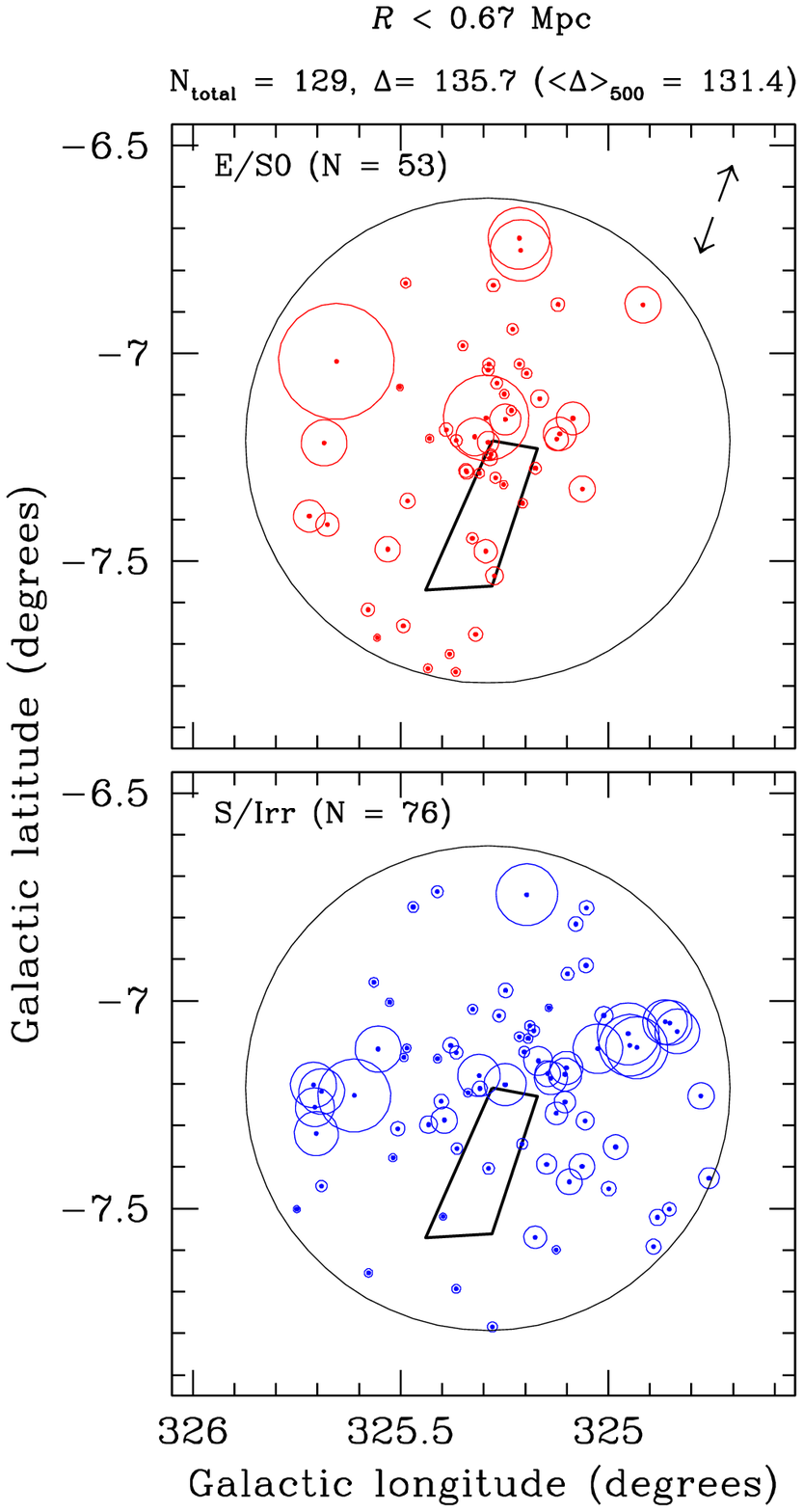,width=5.8cm}} \hbox{\psfig{figure=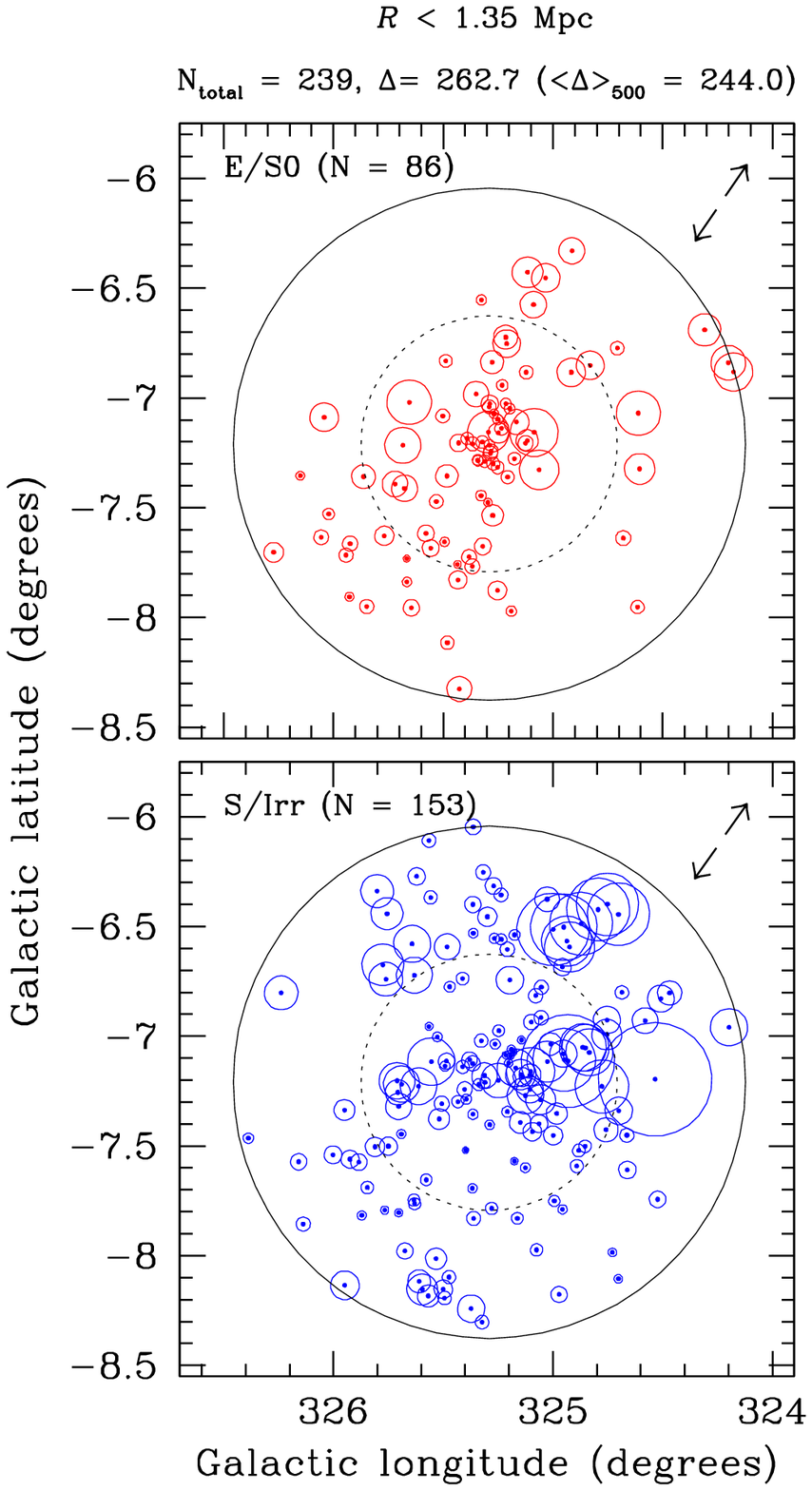,width=5.8cm}}
\hbox{\psfig{figure=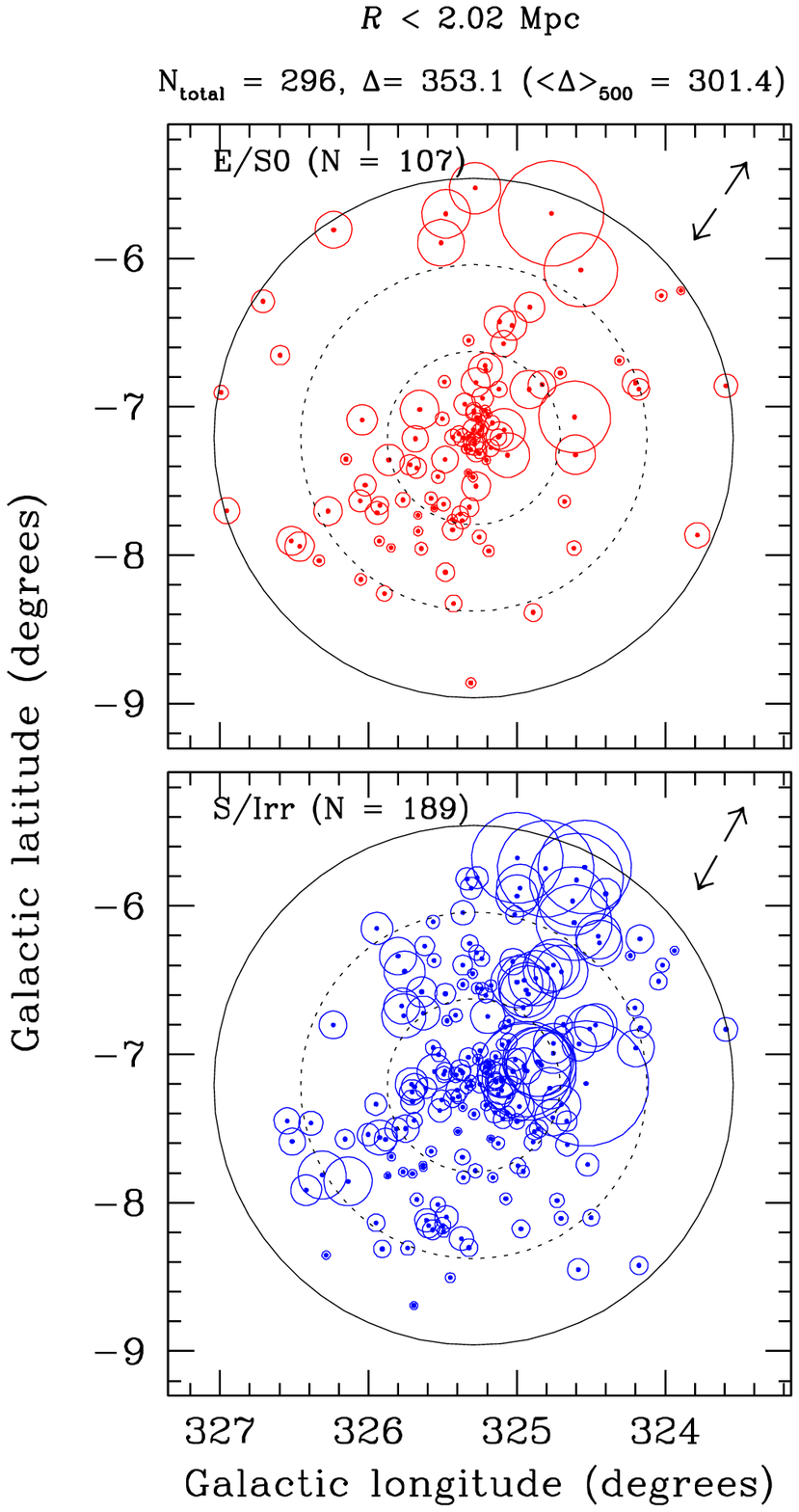,width=5.8cm}}}
 \caption{Results from the Dressler-Shectman $\delta$-test: The distribution in Galactic coordinates of galaxies and their
measured $\delta_i$. The left panels show the results for the inner 0.67 Mpc 
(upper panel: E/S0 galaxies, lower panel: S/Irr galaxies), the middle panels show the results for the inner 
1.35 Mpc (again separated by morphological classification), and the right panels show the results within the 
entire Abell radius (2.02 Mpc). The symbol sizes are proportional to the value of $e^{\delta_i}$, where large 
circles indicate significant deviations from the local mean velocity or local mean velocity dispersion. In the left
panels, the location of the X-ray subcluster (B\"ohringer et al.~1996) is marked by the solid lines. 
If present, the arrows in the top right-hand corner of the panels 
indicate the direction of the position angle of the distribution if significant elongation is detected.}
 \label{dsall}
\end{figure*}

Of the 500 MC simulations,
38\%, 75\% and 14\% revealed a higher $\Delta_{\rm MC}$ than the observed $\Delta$ for the combined, E/S0 and S/Irr 
sample, respectively. Nominally, substructure is said to be present in the observed sample when less than 10\% of 
the MC simulations show a larger amount of substructure. In this case, the S/Irr galaxies
reveal marginal evidence for substructure.
Fig.~\ref{dsall} shows the results of the Dressler-Shectman $\delta$-test for the combined
sample, plotted in Galactic coordinates (as in Fig.~\ref{distribution}). 
We have plotted the E/S0 galaxies (upper-left panel) and the S/Irr galaxies (lower-left panel) separately.
The size of the symbols is proportional to the individual values of $\delta_i$, where large circles indicate
significant deviations from either the local mean velocity or the local mean velocity dispersion. 

Various individual galaxies exhibit properties which are also strongly aligned along the same
position angle (WKK\,6305 = PKS\,1610-605: 
extended head-tail radio continuum emission (Jones \& McAdam 1992), WKK\,6176: extended X-ray tail (Sun et al.~2006)); 
there is an obvious elongation 
within the Norma cluster which affects the way in which the
galaxies interact with the intracluster medium. In Sect.~4 we will review some of the properties of 
these galaxies in more detail.

\subsubsection{The inner $\frac{2}{3}$ Abell radius (= 1.35 Mpc)}

The inner $\frac{2}{3} R_A$ of the Norma cluster contains 239 confirmed cluster members, 
of which 86 are E/S0 galaxies (36\%) and 153 are S/Irr galaxies (64\%). The velocity distribution 
(histogram) is formally consistent with being Gaussian (see middle panel of Fig.~\ref{vhist}), although a slight
excess of spiral galaxies at lower velocities is present.
The velocity centroid of the combined sample ($C_{\rm BI} = 4822 \pm 66$ km s$^{-1}$) is somewhat larger compared 
to that of the inner $\frac{1}{3} R_A$-sample, but this is largely due to an increase in the velocity
centroid of the S/Irr population (although the velocity scale of the S/Irr galaxies has not changed). 
The values for the biweight velocity centroids and scales of the various samples are given in Table~\ref{norsubstr}.

The offset in the biweight velocity centroid between the E/S0 and S/Irr sample remains albeit slightly lower
and is 222 km s$^{-1}$ with a significance of $S_{\rm V} = 1.8$. The velocity centroid of the E/S0 galaxies has not 
changed by extending the sample to a larger radius (see Table~\ref{norsubstr}), although its biweight velocity
scale is somewhat smaller; $S_{\rm BI}$ = 901 km s$^{-1}$ for $R < \frac{2}{3} R_A$, compared to 
$S_{\rm BI}$ = 964 km s$^{-1}$ for $R < \frac{1}{3} R_A$.

In terms of their spatial distribution, the elliptical and spiral populations now both reveal significant elongation and have 
position angles of 102$^{\circ}$ and 101$^{\circ}$, respectively, in the equatorial coordinate frame. This corresponds
to position angles of 146$^{\circ}$ and 145$^{\circ}$ in the Galactic coordinate frame. The latter are again indicated by
arrows in the top-right of the middle panels in Fig.~\ref{dsall}. The $\delta$-test now clearly reveals substructure
in the S/Irr sample (only 8\% of the MC simulations show a larger degree of substructure). Interestingly, the 
$\delta$-test shows that the E/SO population is completely free of any detectable substructure. In Fig.~\ref{dsall},
the results from the $\delta$-test of the combined sample out to $R < \frac{2}{3} R_A$ is shown in the middle panels,
where the upper-middle panel shows the E/SO galaxies and the lower-middle panels displays the S/Irr galaxies. 

\subsubsection{The Abell radius (= 2.02 Mpc)}

Our final sample extends out to the full Abell radius of the Norma cluster. Within this region, there are 296 cluster members
(Sect.~3.1) of which 107 are classified E/S0 (36\%) and 189 belong to the S/Irr population (64\%). 
The velocity histogram of the combined 
set (the hashed histogram in the lower panel of Fig.~\ref{vhist}) shows some evidence for kurtosis, based on the average of
6 kurtosis tests (Pinkney et al.~1996), and a clear excess of galaxies at lower velocities. The difference in the velocity
centroid of the two morphologically-distinct samples is reduced to 167 km s$^{-1}$ (164 km s$^{-1}$ in the cluster rest
frame) at a significance of $S_{\rm V} = 1.5$.

The spatial distribution of both the E/S0 galaxies and the S/Irr galaxies is strongly elongated with position angles
of 102$^{\circ}$ and 107$^{\circ}$, respectively, in the equatorial coordinate frame. This corresponds to 146$^{\circ}$ and
151$^{\circ}$ in the Galactic coordinate reference frame; these angles are indicated by the arrows in the top-right corner of
the right panels in Fig.~\ref{dsall}. The uncertainty in the position angle is $\sim$7$^{\circ}$.

As before, the E/S0 population appears relaxed. Their velocity centroid remains constant throughout the cluster (4979 $\pm$ 85 km
s$^{-1}$ for all the E/S0 galaxies within the Abell radius) and no substructure is detected by the Dressler-Shectman $\delta$-test. 
The velocity scale ($S_{\rm BI}$) of the E/S0 sample shows a distinct decrease as a function of radius (see Table~\ref{norsubstr}), 
again a signature of a relaxed rich cluster (Rines et al.~2003). 

The spiral galaxy population, on the other hand, appears far from relaxed. The velocity centroid increases with increasing radius
(shifting by $\sim$200 km s$^{-1}$ across the Abell radius) and the velocity scale stays roughly constant at
$\sim$960 km s$^{-1}$. The Monte Carlo simulations of the 
Dressler-Shectman $\delta$-test of the S/Irr galaxies indicate that only 1\% of the simulations show a larger degree of substructure
compared to the observed amount of substructure. The results of the $\delta$-test of the combined sample ($N = 296$) is shown in
the right panels of Fig.~\ref{dsall}.

\subsection{Subgroups in the Norma cluster}

We have identified two spiral-rich subgroups based on the Dressler-Shectman $\delta$-test of the S/Irr population alone. When
displaying those S/Irr galaxies for which $\delta_i > 2.25$, two distinct groups appear. In Fig.~\ref{subgroup} the distribution
in Galactic coordinates of the 296 cluster members are shown, where spiral galaxies with $\delta_i > 2.25$ are shown as encircled 
dots.  This $\delta_i$ limit was chosen based on the outcome of the $\delta$-test of the E/S0 galaxies; that sample is completely 
free of any substructure and there the largest measured $\delta_i$ was 2.25. A few isolated galaxies also appear with large 
$\delta_i$ values. 

In the case of WKK\,6406 at ($\ell, b, v$) =  ($325.69^{\circ}$, $-7.22^{\circ}$, $7349 \pm 35$ km s$^{-1}$), its large heliocentric
velocity could indicate that it is a background galaxy which was mistakenly identified as a cluster member (see also Fig.~\ref{raddist}).

Close to the centre of the Norma cluster is a compact group (dubbed `Norma A') where we
have isolated a group of five dynamically distinct galaxies around WKK\,6078 (including WKK\,6071, WKK\,6078, WKK\,6125, 
WKK\,6135 and ZOA\,J16113352). This group is marked in Fig.~\ref{subgroup} by the small solid circle within the $R < \frac{1}{3} R_A$
region (inner dashed circle). The centre of Norma A is approximately at right ascension and declination $16^h12^m00^s$, $-61^{\circ}04'40''$
(J2000.0). Based on these five galaxies, we find a mean velocity of 4453 km s$^{-1}$ (which is 418 km s$^{-1}$ less 
than the mean of the cluster, corresponding to 411 km s$^{-1}$ in the cluster rest frame). Norma A has a velocity dispersion of 
312 km s$^{-1}$, which is much smaller than the velocity scale of the cluster (925 km s$^{-1}$). 

A second dynamically distinct group of galaxies (`Norma B') is found further from the core of the cluster, centred 
around WKK\,5751 (other galaxies include WKK\,5718, WKK\,5779, WKK\,5783, WKK\,5796 and WKK\,5813). This group is indicated by
the large solid circle in Fig.~\ref{subgroup} in the region $\frac{2}{3} R_A < R < R_A$ and has a central position
(in right ascension and declination) of $16^h03^m56^s$, $-60^{\circ}26'54''$ (J2000.0). It has a mean velocity
of 5313 km s$^{-1}$ (an offset of +435 km s$^{-1}$ in the cluster rest frame) and a velocity dispersion of 604 km s$^{-1}$. 

Norma B (and to a lesser extent Norma A) lies along the Norma wall elongation, supporting the idea that cluster infall
occurs along the connecting filaments and wall-like structures. This is consistent with the large-scale structure formation
and evolution as seen in the $\Lambda$-CDM Millenium simulation (Springel et al.~2005).
To gauge how far the merger of both Norma A and B with the main cluster (Norma major) has progressed, deep 
observations with the Australian Telescope Compact Array (ATCA) could be used to determine whether the spiral galaxies 
in Norma A and Norma B are hydrogen-deficient as a result of interactions with the intracluster medium. Previous observations
with ATCA of the Norma cluster (Vollmer et al.~2001) showed that spirals in the Norma cluster are generally 
HI-deficient, but these observations did not include Norma A and B, respectively.

\begin{figure}
\centerline{\hbox{\psfig{figure=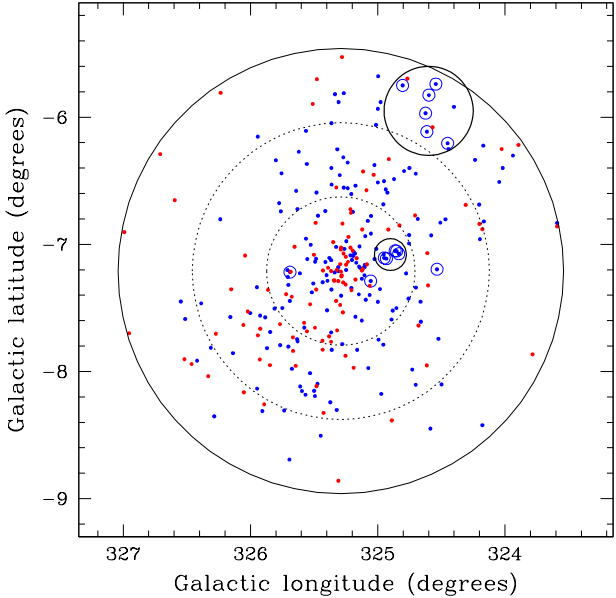,width=8.2cm}}}
 \caption{The distribution in Galactic coordinates of the 296 cluster members within the Abell radius. Encircled dots represent
spiral galaxies with $\delta_i \ge 2.25$ (based on the $\delta$-test of the entire S/Irr galaxy population). Two spiral-rich subgroups
have been identified and are marked by the solid circles.}
 \label{subgroup}
\end{figure}

\subsection{The peculiar velocity of the central cD galaxy}

The Norma cluster contains two large cD galaxies, namely WKK\,6312 at $v_{\rm cD} = 3839 \pm 38$ km s$^{-1}$ 
(Woudt et al.~2004) and WKK\,6269 at $v_{\rm cD} = 5441 \pm 52$ km s$^{-1}$ (Woudt et al.~2004). The latter has
been observed with the 2dF and was found to be in excellent agreement with previous measurements: 
$v_{\rm cD} = 5448 \pm 35$ km s$^{-1}$, see Table~\ref{2dftable}). 
WKK\,6269 is also known as PKS\,B1610-608 (one of the 20 strongest extragalactic
radio sources) and is a textbook example of a wide-angle-tail (WAT) radio galaxy (Jones \& McAdam 1992, 1996). 
Such WAT morphology either reflects the motion of the cD galaxy through the cluster and its interaction with
the intracluster medium via ram pressure (Owen \& Rudnick 1976), or indicates the presence of a 
cluster-subcluster merger (Burns 1998).

Whether WKK\,6269 is at rest with respect to the potential well of the 
cluster can be assessed from its peculiar velocity, i.e.~the difference 
between the velocity centroid of the cluster ($C_{\rm BI}$) and the velocity of the individual cD galaxy ($v_{\rm cD}$). 
The peculiar velocity has to be corrected by a factor $(1+z)$ to ensure the velocity difference is in the cluster
rest frame and considered independently for the varying values in the three regarded spheres.

In Table~\ref{norsubstr} we also list the values of the peculiar velocity of WKK\,6269 as determined within 
the various spheres. It ranges from $\sim$650 km s$^{-1}$ within a radius of 0.67 Mpc to $\sim$550 
km s$^{-1}$ within 2.02 Mpc. The statistical significance of this peculiar velocity ranges from 6.5 to 7.5
($S_{\rm V} \equiv {|v_{\rm cD} - C_{\rm BI}|} / {\sqrt{\sigma_1^2 + \sigma_2^2}}$, where $\sigma_1$ is the error
in the velocity centroid of the cluster and $\sigma_2$ is the error in the velocity measurement of the
cD galaxy). The large number of cluster members used to determine the velocity 
centroid ($\sigma_1 = S_{\rm BI} / \sqrt{N_{\rm gal}}$) and the small error in the measurement of the 
heliocentric velocity of WKK\,6269, makes this large peculiar velocity highly significant. 
It is exceptionally large when compared to other
cD galaxies in clusters (Oegerle \& Hill 2001), but not without precedent (Pimbblet et al.~2006). 
Note that the other cD galaxy in the Norma cluster (WKK\,6312) has an even larger velocity offset.

B\"ohringer et al.~(1996) identified an X-ray subgroup close the centre of the Norma cluster (see Fig.~\ref{optxray}).
This subgroup (dubbed `Norma minor') is fairly massive; Tamura et al.~(1998) estimate that the mass of this subgroup could add 
up to $\sim$50\% to the total mass of the cluster.

A comparison of the 843 MHz radio continuum emission of PKS\,B1610-608 (Jones \& McAdam 1992) with 
the X-ray contours of this central subgroup (shown in Fig.~\ref{central}) 
shows that the radio lobes of WKK\,6269 are closely aligned 
with the X-ray subgroup. The large observed peculiar velocity of the cD galaxy in the Norma cluster
is most likely caused by this ongoing merger. 

Based on the compactness of the X-ray subcluster, B\"ohringer et al.~(1996) 
argued that the merger has not progressed very far yet, and that most of the main component of the cluster 
is still undisturbed by the collision. This is consistent with simulations of cluster mergers (Pinkney et al.~1996), 
which show that large peculiar velocities can be reproduced in the event of large-scale mergers at the time of 
core-crossing. If this merger takes place close to the plane of the sky, it would also explain the non-results 
of the statistical tests. These are least sensitive to mergers occuring perpendicular to the line-of-sight.
Therefore, the X-ray morphology -- in combination with the large peculiar motion of the central cD 
galaxy -- strongly suggests a recent or commencing merger at the core of the cluster.

\subsection{Dynamical mass estimate}

For the determination of the dynamical mass of the Norma cluster, we have used both the 
virial theorem ($M_{\rm VT}$) and the projected mass estimator ($M_{\rm PME}$), see equations 21 and 22 of Pinkney et al.~(1996).
The use of the bi-weight velocity centroid and scale (Beers et al.~1990) in the virial theorem (instead of the velocity
mean and standard deviation) leads to a more robust mass estimate ($M_{\rm RVT}$). The latter is more robust against the effects of contamination
by the inclusion of possible non-members in the analysis. The projected mass estimator (Bird 1995), on the other hand, is sensitive to the 
presence of (spatially-separated) subclusters due to its proportionality to the projected distance between galaxy $i$ and the cluster
centroid ($R_{{\perp},i}$) (see equation 22 in Pinkney et al.~1996). The presence of a spatially-separated subcluster (e.g. in a premerger configuration)
would result in a systematic offset with respect to the cluster centroid; this leads to larger values of $R_{{\perp},i}$ and thus to a significantly 
larger mass estimate.
For a full discussion of the appropriate use of these dynamical mass estimators we refer to Pinkney et al.~(1996) and Bird (1995).

The three dynamical mass estimates ($M_{\rm VT}$, $M_{\rm RVT}$ and $M_{\rm PME}$)
determined within the three radial limits (using the combined samples of $N$ = 129, 239 and 296
galaxies, respectively) are given in Table~\ref{massnorma}. On average, $M_{\rm RVT}$ is $\sim 5$\% larger than $M_{\rm VT}$. The projected
mass estimate, however, is generally about 50\% larger than $M_{\rm VT}$ and indicates the presence of a spatially-distinct subcluster
(projected on the plane of the sky) presumably in the early stages of merging (Pinkney et al.~1996). This is consistent with our 
previous indications of subclustering, particularly in the form of Norma minor (the X-ray subgroup).

B\"ohringer et al.~(1996) and Tamura et al.~(1998) both give an estimate of the gravitational mass of the 
Norma cluster based on ROSAT and ASCA X-ray observations, respectively.  In Table~\ref{massnorma}, we list the values (converted from $h_{50}^{-1}$
to $h_{73}^{-1}$) of the mass within a specific radius as derived from X-ray observations by B\"ohringer et al.~(1996) and Tamura et al.~(1998).
Both virial mass estimates ($M_{\rm VT}$ and $M_{\rm RVT}$) are consistent with the mass determined from the X-ray luminosity of the cluster.
In the presence of substantial subclustering, as suggested here for the Norma cluster, all dynamical mass estimators could still overestimate the
true mass of the cluster, depending on the projection angle of the cluster--subcluster merger axis with respect to the line of 
sight (Pinkney et al.~1996). This effect is smallest for $M_{\rm VT}$ and $M_{\rm RVT}$ for a merger occuring perpendicular to the line-of-sight.
We can therefore safely conclude that the mass of 
the Norma cluster within the Abell radius corresponds to $1-1.1 \times 10^{15} h_{73}^{-1}$ M$_{\odot}$. 
This confirms the status of the Norma cluster as the most massive cluster in the Great Attractor.

\begin{table}
 \centering
  \caption{Mass estimates of the Norma cluster.}
  \begin{tabular}{@{}lcl@{}}
\hline
  \multicolumn{3}{c}{X-ray mass (gravitational)}\\[2pt]
  $R$ (h$_{73}^{-1}$ Mpc)$^*$     & $M(<R)$ (h$_{73}^{-1}$ M$_{\odot}$)$^*$  &  Reference \\
\hline
   0.68 & \hfill $1.5 - 4.0 \times 10^{14}$  & B\"ohringer et al.~(1996) \\
   0.75 & \hfill $3 \times 10^{14}$          & Tamura et al.~(1998) \\
   2.05 & \hfill $2.9 - 15 \times 10^{14}$   & B\"ohringer et al.~(1996)\\
\hline
  \end{tabular}
  \begin{tabular}{@{}lccc@{}}
  \multicolumn{4}{c}{Dynamical mass}\\[2pt]
  $R$ (h$_{73}^{-1}$ Mpc) \hspace{0.8cm}    & \multicolumn{3}{c}{$M(<R)$ (h$_{73}^{-1}$ M$_{\odot}$)}\\
  all galaxies               & $M_{\rm VT}$  &  $M_{\rm RVT}$  & $M_{\rm PME}$ \\
\hline
   0.67   & \hfill $4.2 \times 10^{14}$ & \hfill $4.3 \times 10^{14}$  & \hfill $6.6 \times 10^{14}$ \\
   1.35   & \hfill $8.1 \times 10^{14}$ & \hfill $8.6 \times 10^{14}$    & \hfill $11.6 \times 10^{14}$ \\
   2.02   & \hfill $10.4 \times 10^{14}$& \hfill $11.0 \times 10^{14}$   & \hfill $14.7 \times 10^{14}$ \\
\hline
\end{tabular}
\label{massnorma}
{\newline \footnotesize{$^*$ The original values have been converted from $h_{50}^{-1}$ to $h_{73}^{-1}$. \hfill }}
\end{table}

\section{Individual galaxies}

A number of galaxies in the Norma cluster show direct or indirect evidence of interaction with 
the intracluster medium (ICM). Here, we will explore these galaxies in some 
detail in the light of the preceding discussion.


%
\subsection{WKK\,6176 and the X-ray tail}

Recent {\it Chandra} and {\it XMM-Newton} observations of WKK 6176 (= ESO 137-001) 
revealed the presence of a $\sim$70 kpc long X-ray tail pointing away from the
cluster centre (Sun et al.~2006), suggesting this galaxy is undergoing a significant amount 
of gas stripping. The extent of this X-ray tail is unusual (Sun et al.~2006).


We have deep $B$, $V$ and $R_C$-band photometry of WKK\,6176. These data were obtained in 1999 May with the MPG/ESO 2.2-m telescope
at la Silla and the Wide Field Imager (ESO Programme 63.N-0054). We covered the entire Abell radius of the Norma cluster
for the purpose of measuring the $R_C$-band luminosity function (see also Fig.~\ref{optxray}). This optical 
as well as the near-infrared $J$, $H$ and $K_s$ luminosity function of the Norma cluster will be 
presented in a separate paper in this series.

\begin{figure}
\centerline{\hbox{\psfig{figure=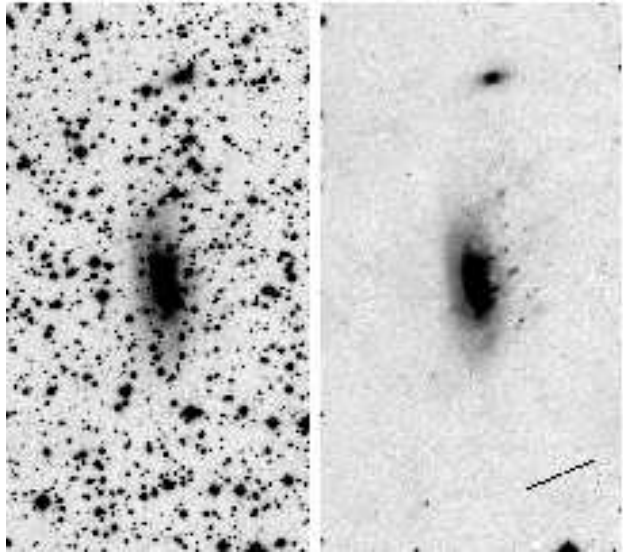,width=8.2cm}}}
 \caption{The $R_C$-band image of WKK6176 before (left panel) and after star subtraction (right panel) using
the {\tt KILLALL} routine (Buta \& McCall 1999) in IRAF. The field of view is $2.2 \times 4.0$ arcminutes, 
North is up and East is to the left. The diagonal bar at the bottom-right in the star-subtracted image indicates
the position angle of the elongated E/S0 population within $R < 0.67$ Mpc.}
 \label{wkk6176r}
\end{figure}

WKK\,6176 is located close to the core of the Norma cluster. In Fig.~\ref{optxray} it can be found at (RA, DEC) $\sim$
(16.22425$^{\circ}$, --60.76397$^{\circ}$), to the lower-left of the right-most X-ray contours (aligned with 
the virtual extension of the X-ray 
subgroup). A close-up of WKK\,6176 is shown in Fig.~\ref{wkk6176r} where we have displayed the $R_C$-band image
in a field of view of $2.2 \times 4.0$ arcminutes.
Given the low Galactic latitude of the Norma cluster ($b \sim -7^{\circ}$) and proximity to the 
Galactic bulge (only 35$^{\circ}$ away), a large number of stars are superimposed on the galaxy
images. Reliable photometry can only be obtained after careful
removal of the many foreground stars. We used the {\tt KILLALL} routine developed by Buta \&
McCall (1999) within the IRAF\footnote{IRAF is distributed by the National Optical Astronomy Observatories,
which are operated by the Association of Universities for Research
in Astronomy, Inc., under cooperative agreement with the National
Science Foundation.}  environment to remove this stellar contamination. 
Fig.~\ref{wkk6176r} illustrates the effectiveness of this star-removal procedure for the 
Norma cluster galaxy WKK\,6176 (compare the right-hand panel of Fig.~\ref{wkk6176r} to the original image). It also reveals 
numerous striking low-brightness filaments to the west of WKK\,6176, appearing to stream away from WKK\,6176 at a position
angle of $\sim$125$^{\circ}$. 
Several bright knots (distinctly different from the Galactic foreground pollution) appear within 
these filaments. The low surface brightness filaments are aligned with the X-ray tail (Sun et al.~2006), but are not only
confined to the region of the X-ray tail. For comparison, we have indicated the direction of the major axis of the E/SO
galaxy population with the diagonal marker in Fig.~\ref{wkk6176r}.

WKK\,6176 is a low-redshift equivalent of the two recently detected spiral galaxies in massive rich clusters 
(Abell 2667 and Abell 1689) at $z \sim 0.2$ which show clear evidence for galaxy transformation (Cortese et al.~2007). 
Interestingly, WKK\,6176 is located at a similar projected distance from the centre of the Norma cluster (0.28 Mpc) as
the two high-redshift spirals in Abell 2667 and Abell 1689, which lie at 0.34 h$_{70}^{-1}$ Mpc and 0.24 h$_{70}^{-1}$ Mpc 
from their respective cluster centre.

A full investigation into the properties of WKK\,6176 as derived from multiwavelength photometry ($B V R_C J H K_s$), spectroscopy,
and galaxy evolution modelling (Fritze-von Alvensleben \& Woudt 2006), and its implications for galaxy evolution in dense environments
will be presented elsewhere.

\subsection{WKK\,6305: the head-tail radio continuum source}

\begin{table}
 \centering
  \caption{Selected galaxies in the Norma cluster.}
  \begin{tabular}{@{}llc@{}}
\hline
 Galaxy  & Observational characteristic & Pos.~angle \\
\hline
 WKK\,6176 & -- 70 kpc X-ray tail & 129$^{\circ}$ \\
           & -- Optical filaments & 125$^{\circ}$ \\
 WKK\,6269$^{\ast}$ & -- Central cD galaxy & 128$^{\circ}$ \\
 WKK\,6305 & -- 500 kpc radio-continuum tail & 108$^{\circ}$ \\
\hline
\end{tabular}
{\footnotesize{\newline $^{\ast}$ The position angle quoted here for WKK\,6269 is a mean \hfill \ 
\newline position angle as determined from GALFIT (Peng et al.~2002) \hfill \ 
\newline isophotal fitting of deep $K_s$-band imaging. \hfill \ }}
\label{interact}
\end{table}

Another peculiar galaxy in the Norma cluster is WKK\,6305, also known as PKS\,1610-605 (Jones \& McAdam 1996). It is located
at a similar distance from the centre of the cluster (0.29 Mpc) as WKK\,6176.
In Fig.~\ref{central} we show an overview of the central region with the same field of view as
Fig.~\ref{optxray}. The galaxy distribution of confirmed cluster members and the X-ray subgroup are as before, but 
now the radio continuum emission of WKK\,6269 and WKK\,6305 at 843 MHz are overplotted (reproduced from Jones \& McAdam 1992). 
WKK\,6305 corresponds to the head-tail source visible in Fig.~\ref{central}. The tail length of 26$'$ ($\sim$ 500 kpc at the
distance of the Norma cluster) represents one of the longest radio continuum tails observed. The position angle of the tail
is $\sim$108$^{\circ}$ (Jones \& McAdam 1996) and is, as before with the X-ray tail of WKK\,6176, closely aligned with the
elongated (E/S0) galaxy distribution in the cluster.

\begin{figure}
\centerline{\hbox{\psfig{figure=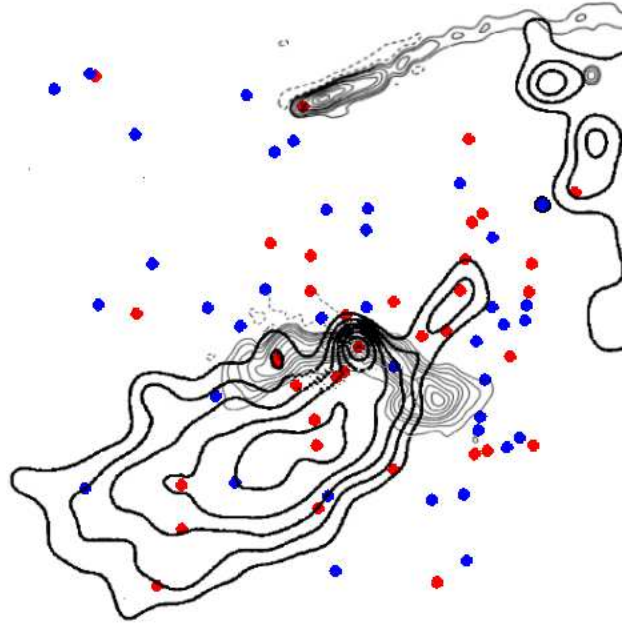,width=8.2cm}}}
 \caption{The galaxy distribution and the X-ray subgroup of the inner $\sim$0.66 Mpc $\times$ 0.66 Mpc
(as in Fig.~\ref{optxray}) with the radio continuum emission (reproduced from Jones \& McAdam 1992) 
of WKK\,6269 and WKK\,6305 overplotted. WKK\,6176 is indicated by the large dot in the virtual 
extension of the X-ray subgroup.}
 \label{central}
\end{figure}

\section{Discussion}

The dynamical analysis of the Norma cluster presented here has revealed a significant amount of subclustering 
in this nearby rich cluster, ranging from the central X-ray group to the two spiral-rich subgroups further from the
core of the cluster. Even though the X-ray group did not show up in the dynamical analysis, the large peculiar
velocity of the cD galaxy, whose radio lobes appear to `embrace' the X-ray contours of this central group, is a tell-tale
sign of an ongoing merger. The large discrepancy between $M_{\rm PME}$ and the mass determined from the virial theorem (where
$M_{\rm MPE} \sim 1.5 \times M_{VT}$) is an independent indication of the presence of a spatially-separated
subcluster of substantial mass.
Significant subclustering is not unusual for rich and massive clusters; Colless \& Dunn (1996) 
in their detailed dynamical analysis of the archetypical rich Coma cluster also revealed the presence of an ongoing merger.

The galaxy distribution in the Norma cluster is clearly elongated, with a position angle (in equatorial coordinates)
ranging between 116$^{\circ}$ (for the central part) and $\sim$105$^{\circ}$ (for the overall distribution). It should be
emphasized that this observed elongation is not an artefact of selective Galactic extinction effects at this low
Galactic latitude. The position angle of the major axis of the Norma cluster, as indicated by the arrows in Fig.~\ref{dsall},
runs nearly perpendicular to the lines of constant Galactic foreground
extinction (compare with Fig.~\ref{distribution}). The elongated galaxy distribution is aligned with the major
large-scale structure in this region as can be seen in Fig.~\ref{lssoverview}. Such an alignment is not unexpected 
within the cluster-rich GA environment (Binggeli 1982). 
 
Within the cluster itself, various galaxies show clear evidence for interactions with the intracluster
medium. An overview of these galaxies and their defining characteristics is given in Table~\ref{interact}.
The defining features of these galaxies are strongly aligned with the general galaxy distribution of the cluster. In the
case of WKK\,6176 and WKK\,6305, they are X-ray/optical and radio continuum tails, respectively, whereas for WKK\,6269 (the central cD galaxy)
the major axis of the galaxy is aligned with the cluster (see Table~\ref{interact}).

\section{Conclusion}

The Norma cluster (ACO 3627) is a nearby, rich and massive cluster -- on par with the more distant Coma cluster -- 
which resides at the bottom of the potential well of the Great Attractor. 
The galaxy distribution of the cluster members shows a clear elongation which is aligned with the main
wall-like structures of the GA. Despite the relaxed appearance of the early-type galaxy population in the
Norma cluster, a large amount of subclustering is present. We have identified two spiral-rich subclusters (Norma A and B) in
addition to the previously identified central (X-ray) subcluster (Norma minor). The ongoing merger of the latter with the main 
cluster (Norma major) is assumed to be responsible for the large peculiar motion of the central cD galaxy. 

The proximity of the Norma cluster offers an excellent opportunity to study the interaction of cluster members such as WKK\,6176
with the intracluster medium at high resolution and sensitivity.

\section*{Acknowledgments}

We thank J.~Pinkney for providing his cluster substructure analysis programme and M.~McCall for the
use of his KILLALL routine.
This research has made use of the NASA/IPAC Extragalactic Database (NED) which is 
operated by the Jet Propulsion Laboratory, California Institute of Technology, under 
contract with the National Aeronautics and Space Administration. PAW, RCKK and APF kindly acknowledge
funding from the National Research Foundation.

\end{document}